\documentstyle[preprint,tighten,prd,aps]{revtex}

\begin{document} \draft 
\preprint{INFN PI/AE 99/08}
\title{A stronger classical definition of
Confidence Limits} \author{Giovanni Punzi} \address{Scuola Normale
Superiore and INFN, Sezione di Pisa\\
P. dei Cavalieri 7, I-56100 Pisa,
Italy} \date{\today} \maketitle

\begin{abstract}

A novel way of defining limits in classical statistics is proposed. This
is a natural extension of the original Neyman's method, and has the
desirable property that only information relevant to the problem is used
in making statistical inferences. The result is a strong restriction on
the allowed confidence bands, excluding in full generality pathologies
as empty confidence regions or unstable solutions. The method is
completely general and directly applicable to all problems of limits.

\end{abstract}


\section{Introduction} \label{sec:Intro}

The concept of Confidence Region for a parameter at a given Confidence
Level is a center piece of classical statistics and was first introduced
by Neyman\cite{Neyman}. It gives a definite meaning to the making of
statistical inferences about the region where the value of an unknown
parameter might fall, without any assumption on whether the parameter
can be attributed some probability distribution and what it might be. An
alternative approach to setting acceptance regions for a parameter is
from Bayes, that on the opposite assumes and explicitly incorporates the
information from a probability distribution of the parameter, which is
supposed to be known before the measurement of the data set in hand, and
it is therefore called ``a priori" distribution.

With regard to the choice between the two methods of statistical
inference, the Author shares a common opinion that Bayesian methods are
very useful whenever there is a solid ground for establishing the "a
priori" parameter distribution, which this method readily exploits in
optimal way, but the classical methods are the only reasonable choice
whenever this does not happen. Unfortunately the measurements of physics
quantities belong almost always to the second class. The widespread
preference of the physicists for classical methods of setting acceptance
regions seemed recently to weaken when it was realized that the usual
procedures for setting limits in the classical framework can lead in
some cases to highly counter-intuitive results.

Several solutions to this unpleasant situation have been proposed, some
of them requiring partial fallback on Bayesian
concepts\cite{Giunti,PDG}, or even argued that the classical method was
fundamentally weak, and could not work without the supplement of some
Bayesian ingredient.

Other authors \cite{F-C,Roe-Wood}, defended the classical point of view
by proposing some alternative methods for setting limits that eliminate
the unpleasant results while still adhering to Neyman's prescription.
The present work follows that same line of looking for meaningful
results within the classical approach, avoiding any Bayesian
contamination. However, I argue that none of the previous proposals is
completely satisfying, and that a deeper revision of current ideas is
needed in order to really solve the difficulties, yielding to very
different conclusions from past work on the subject.

It is worth noting that the insistence on classical methods should not
be taken to imply that the Bayesian method are not very useful in the
more limited field where they are unambiguously applicable.

In Sec.\ \ref{sec:problems} a few examples of problematic limits are
discussed, some of which appear not to have been previously considered.
In Sec.\ \ref{sec:analysis} I analyze the reasons for the physicist's
dissatisfaction and what they reveal about the incompleteness of the
classical CL definition by Neyman, and in sec.\ \ref{sec:proposal} I
propose a general solution of these issues completely contained in the
realm of classical statistics. In Sec.\ \ref{sec:properties} the most
important features of the proposed approach are discussed, with brief
notes on some specific examples.

\section{Problems with standard classical limits}\label{sec:problems}

\subsection{Definitions and notations}\label{sec:definitions}

Let $\mu\in M$ indicate some unknown parameters, and $x\in X$ a random
variable we can observe, whose probability distribution $p(x|\mu)$ ({\em pdf} for short)
depends in some way on the unknowns $\mu$. Both $\mu$ and $x$ can be
arbitrary objects, e.g. they can be vectors of real numbers of any
length. When the observable is continuous a probability {\em density}
rather that a discrete distribution is necessary to describe it, but for
simplicity the same notation $p(x|\mu)$ will be used, and the
distinction will be explicitly noted only when necessary. In both cases
$p(x\in S|\mu)$ will indicate the total probability for the observable
to fall in a given subset $S\subset X$, independently on whether it is
obtained by a sum (discrete variable) or an integration (continuous), or
both\footnote{Note that $p(x\in \{\bar{x}\}|\mu)=p(\bar{x}|\mu)$ for discrete variables, 
while $p(x\in \{\bar{x}\}|\mu)=0$ for continuous variables, independently
of the value of the density $p(\bar{x}|\mu)$ at the point $\bar{x}$.}.

Let $B(x)$ be {\em any} function associating to each possible observed
value of $x$ a subset of values of $\mu$ ($B$ is intended to represent
some algorithm to select ``plausible" values of the unknown $\mu$ on the
basis of our observation). The classical definition of CL from Neyman
can then be stated as follows:  the function $B$ (``confidence band") is
said to have ``Confidence Level" equal to $CL$ if, whatever the value of
$\mu$, the probability of obtaining a value of $x$ such that $\mu$ is
included in the accepted region $B(x)$ is (at least) CL. In short:

\begin{equation}\label{eq:Neyman} CL(B) = \inf_\mu\ p(\mu\in B(x)|\mu) = 
1 - \sup_\mu p(\mu \not \in B(x)|\mu)
\end{equation}

Obviously the Confidence Level is a property of the band $B$ as a whole,
not of a confidence region associated to a particular value of $x$: it
is quite possible for two different algorithms $B$ and $B'$, to give the
same confidence region for some $\bar x$, and still have very different
Confidence Levels. This is the reason for the need of always deciding
the algorithm $B$ before making the actual measurement, clearly implied
by the original formulation, but apparently often forgotten, and only
recently clearly pinpointed\cite{F-C}.

Neyman's definition is so general, that after choosing the desired CL,
there is a very wide variety of bands $B$ satisfying it. In a generic
case, confidence regions can be arbitrarily complicated subsets of the
$\mu$ space. One can even construct fractal confidence region if one
likes to do so.

For this reason, soon some ``rules" have been invented to easily obtain
simple confidence regions with desirable properties. Most of them are
based on ordering all possible values of the observable $x$ according to
some rule, and then determining the confidence region by adding up in
order as many values as needed for reaching the desired {\em coverage},
that is the integral of the pdf over the accepted region. Common
examples of rules are upper/lower limits, based on ordering for
increasing/decreasing value of $x$ (assumed a number), ``centered" (for
unidimensional $x$, order by decreasing tail probability, yields equal
probabilities in the upper and lower excluded regions), and the band
obtained by ordering for decreasing $p(x|\mu)$\cite{Crow} (``narrowest
band", or ``Crow band" in the following).

These rules really have nothing fundamental, but they have been so
commonly used that they have been sometimes identified with the very
essence of the CL concept. For this reason, when some examples were
found that showed serious limitations of these rules, their failure has
been sometimes perceived as a failure of classical statistics as a
whole, and alternative solutions often looked for in Bayesian concepts.

Obviously, other choices can be singled out within the huge space of
classical solutions, to give satisfactory solutions to those cases. In
order to overcome the limitations of the other methods, the new method
of Likelihood Ratio (LR) ordering\footnote{This is often referred to in
the literature as ``unified approach" due to its capability of producing
a single band containing both ``central" and ``upper/lower" intervals,
but that property is not of particular relevance in the present context,
therefore the more explicit expression LR--ordering is adopted} has
been recently proposed\cite{F-C}. This amounts to order the observable
values by decreasing $p(x|\mu)/p(x|\hat\mu)$, where $\hat\mu$ represents
the maximum likelihood estimate of $\mu$, given $x$. This method appears
to have distinct advantages over the previous, and stirred great
interest around this problem. However, it does have limitations, that
have inspired some amendments\cite{Giunti,Roe-Wood}.

I will argue in the next subsection that the LR--ordering method and its
modifications have pitfalls as serious as those of other methods they
are intended to replace, and cannot therefore be considered a genuine
solution.

\subsection{Specific examples}\label{sec:examples}

I proceed now to examine some examples of problematic confidence bands.

The pathologies encountered are essentially of two kinds. The first and
more obvious is when the confidence region happens to be the empty set.
I avoid to speak of ``unphysical values" of the parameters because I
find it a confusing terminology: in every problem the parameters can
assume values inside some domain, determined by the nature of the
problem. If that domain actually describes all conceivable values of
parameters for which a $p(x|\mu)$ exists, then there is no meaning in
referring to hypothetical values outside that domain: they just do not
exist as possible values for $\mu$. On the other hand, if the
formulation of the physical problem allows to attach a meaning to other
values of the parameters, they should be taken into account from the
start, and cannot be called ``unphysical". Similar considerations apply
to the expression ``the maximum of the likelihood function lies outside
the physical region": the expression usually really means that the
maximum occurs on the border of the parameter space, which does not
poses particular problems and certainly does not suggest arbitrary
extrapolations of the likelihood function outside its domain of
existence.

The other possible pathology is to have ``unreasonably small" confidence
regions, that is actually just a softer version of the previous. It is
less obvious to detect, but it should be clear that it is just as
unacceptable from the physicist's point of view. Also, it is potentially
more dangerous since the experiment result will superficially appear to
convey a great deal of information. How do we {\em know} that a limit is
too tight ? A possible symptom of this situation is when the limits
become tighter with decreasing experiment sensitivity, as in the example
of Poisson with background below.

\subsubsection{Poisson with background: a sensitivity paradox}

Let us examine briefly this problem of Confidence Limits of great
practical importance. The probability distribution is given by:

\begin{equation} p_b(n|\mu) = e^{-(\mu+b)} {(\mu+b)^n \over n!}
\end{equation}

While the observed number of counts $n$ can only be positive, the
presence of a background $b$ constraints the overall mean $\mu + b$ of
the Poisson to be larger than $b$, and therefore creates the possibility
of ``negative fluctuations" in the form of occurrence of much less
observed counts than the average level of background. The``usual"
ordering rules mentioned in sec.\ \ref{sec:definitions} readily produce
empty confidence regions in that case.

The LR--ordering prevents this, but its results are counter-intuitive
and hard-to-interpret as well.

The problem appears clearly when comparing the results of experiments
observing the same number of counts, but affected by different levels of
background. It is easy to see that with the LR method the upper limit on
$\mu$ goes to zero for every $n$ as $b$ goes to infinity, so that a low
fluctuation of the background entitles to claim a very stringent limit
on the signal. This means that the limit can be much more stringent than
in the case of zero observed events and zero background. This is clearly
hard to accept.

The modification proposed in \cite{Giunti} only softens this behavior,
and in addition uses Bayesian concepts in its formulation, therefore the
uncompromising classical physicist will not want to consider it.

The absurdity of the result is best seen by looking at the case of zero
observed events. This has been clearly pointed out in\cite{Roe-Wood}.

If there is no background, and one observes zero events, one knows that
no signal event showed in the sample at hand, and one can deduce an
upper limit on $\mu$ from this fact. If there is some level of
background and one observes zero events, that implies two facts:

\begin{itemize} \item{a)} no signal event showed in the current sample
\item{b)} no background event showed in the current sample \end{itemize}

The two occurrences are statistically independent, by assumption of
Poisson distribution, therefore they can be considered separately. Fact
b) is totally uninteresting for what concerns the signal: our only help
in making decisions about $\mu$ is fact a), which is exactly the same
information we had in the case of no background. A sensible algorithm
{\em must} therefore give the same upper limit on $\mu$ in the
zero-count case, {\em whatever} the expected background.

This failure is particularly important if one considers that this
behavior stems from the same root as the other problem that the LR
proposal is intended to cure. In fact, the problem can be summarized by
saying that the low likelihood of occurrence of event b) "fools" the
algorithm into making up a very narrow confidence region that has no
basis in what we actually learned from the experiment. This is exactly
the same mechanism that leads to empty regions with the older rules: the
rarity of a set of results is taken as a good reason for rejecting
values of the parameter even if it is uncorrelated with the value of
that parameter. This should make us dubious about the question of
whether the approach of LR ordering really addresses the issue.

This problem was not missed by the proponents of the method, who devote
a section of their paper to it \cite{F-C}. They maintain that the
concern for this problem is motivated by ``a misplaced Bayesian
interpretation of classical intervals", but nonetheless suggest that in
this kind of cases the experimentalist should not publish just the
limits, but also an additional quantity to represent the `sensitivity'
of the experiment. This however avoids the question of how to provide an
interval that properly and completely represent the results of the
measurement, including all information about the sensitivity, that is
the question the present work tries to address.

The modification of LR--ordering proposed in \cite{Giunti} to address
this problem is based on Bayesian quantities, therefore the
uncompromising classical physicist will not accept it. Furthermore, it
only softens the problem rather than eliminating it.

A nice classical solution to this dilemma has been presented in Ref.
\cite{Roe-Wood}, based on explicitly eliminating the spurious
information from the calculation of the coverage, while still ordering
the observable values according to LR. The amount of background events
in the sample is forced to be less than the total number of observed
events. This modification removes the paradoxical behavior of the
limits, and produces results which seem reasonable from all points of
view, so the particular problem of the Poisson with background might be
considered as solved.

However, the above procedure appears to be {\em ad hoc}, and it is not
clear how to apply it to different situations, like the other examples
of this section. In addition to that, the example that follow will show
an important weakness of that variant, and {\em any} other variant based
on the LR ordering rule.

\subsubsection{Gaussian with positive mean}

This is another very important example: $p(x|\mu)$ is gaussian, but the
condition $\mu>0$ holds. If one tries to apply the Crow band, which is
the usual choice for the unbounded case, one gets empty confidence
region for $x<-1.96$ at 95\% CL. This does not happen if one uses the LR
ordering rule, as extensively discussed in \cite{F-C}.

This example makes a very good case for the LR method, but unfortunately
it is easy to expose its instability. Consider a modification of the
gaussian {\em pdf} obtained by adding a second, very narrow gaussian of
the same height but negligible width and area. Let the second gaussian
be centered at a different location, for instance $\mu_2=-1/\mu$. What
is important is just that $\mu_2\rightarrow -\infty$ as $\mu\rightarrow
0$.

Intuitively, this is a very small change of the problem: it just means
that in a {\em negligible} fraction of cases the measurement $x$ will
fall in a different, narrowly determined location. This is not so
artificial an example as it may seem, since it is quite possible for an
experimental apparatus to have rare occurrences of singular responses.

How the confidence regions should change, according to  common sense ?
If the probability of this occurrence is very small (let's say $\ll
1-CL$) one would just ignore the possibility and quote the same
confidence limits as before. One would therefore want from a sound
algorithm to yield very similar bands to the unperturbed case.
Unfortunately, this does not happen with the LR ordering method: since
the ordering is based on the value of the {\em maximum} of the pdf, the
narrow peak of negligible physical meaning is capable of altering the
ordering completely: the maximum of the Likelihood is now a constant for
every value of $x$, and the resulting band goes back suddenly to
something very similar to the old Crow band, that is just ordering by
$p(x|\mu)$. For large negative deviations, the intervals are not exactly
empty, but contain a tiny interval centered around the peak of the
second gaussian.  However, this hardly makes the result satisfying from
a physicist's point of view. When observing a large negative deviation,
it is much more likely that is comes from the tail of the main gaussian
rather than from the ``extremely rare" second gaussian, and one would
like the confidence limits to reflect this fact. The response of the LR
method that instead ``completely forgets" the main gaussian to focus on
the secondary peak, no matter how narrow, appears as a crucial failure.
From a practical point of view, this kind of instability of the solution
means that the response of the apparatus must be known with infinite
precision in order to be able to use the algorithm.

Note that the problem is intrinsic to the {\em ordering}, therefore any
modification of the method acting only on the coverage criteria, as the
one proposed in \cite{Roe-Wood} for handling the Poisson case, will be
plagued by the same problem.

It is also worth reflecting on what happens if the second peak is not so
narrow, but rather comparable to the main peak. In that situation, the
LR algorithm might give a result which is not so violently in contrast
with common sense. Yet, it is hard to avoid the suspect that also in
that case the result will be, in some ill--defined way, not what a
physicist wants.

\subsubsection{Empty confidence regions are not ruled out by the LR
method} The previous example showed a case where LR ordering yields
negligibly narrow confidence regions. For completeness, it is worth
noting that it is also possible to formulate examples where the
LR--ordering produces {\em completely empty} confidence regions on wide
ranges of the observable, contrary to what is generally assumed.

This can be obtained, for instance, by adding to the pdf a narrow,
wiggling ridge of ever increasing height, still of negligible area. For
instance, in the previous example one might simply add to the pdf the
function: $$\epsilon N(x_0+ \delta \sin(\mu),{\alpha\over 1+\mu})$$

where $N(m,\sigma)$ stands for the Gaussian function with unit area,
mean $m$ and standard deviation $\sigma$, and $\delta$, $\alpha$, and
$\epsilon$ are real numbers ( $\alpha$ and $\epsilon$ are ``small"). It
is easy to see that the likelihood function for any $x\in[x_0-\delta,
x_0+\delta]$ has periodic ``spikes" with a height that increases without
limit as $\mu\rightarrow \infty$, therefore the maximum likelihood is
infinite, and the LR is zero for all $x\in[x_0-\delta, x_0+\delta]$ and
all values of $\mu$, including the points on the spikes themselves. As a
consequence, all points in the interval $x\in[x_0-\delta, x_0+\delta]$
will get the lowest possible rank in the ordering, so they will be the
last to be added to the accepted region, for all $\mu$. If an interval
is chosen in such a way that $p(x\in[x_0-\delta, x_0+\delta] |\mu)<1-Cl$
 (which is always possible whatever the pdf), then the confidence region
will be the empty set for all $x\in[x_0-\delta, x_0+\delta]$.

The example is clearly very artificial, but is nonetheless valuable in
signaling the existence of a problem.

\subsubsection{Uniform distribution}

An example which is simpler than the previous and totally plausible in
practice, yet presents unexpected difficulties is the uniform
distribution:

\begin{equation} p(x|\mu) = 1 \text{ if }  \mu<x<\mu+1\text{, otherwise
0}. \end{equation}

Let's consider the case of the domain of $\mu$ being the full set of
real numbers. The upper/lower limits presents no trouble in this case,
but both the Crow band and the LR band are indeterminate, since every
value of $x$ gets assigned the same rank, for whatever $\mu$, therefore
{\em any} band satisfying Neyman's condition will satisfy both. In
particular, note that LR--ordering does not exclude empty confidence
intervals in this example. Again this indicates that the root of the
difficulties that motivated this approach has not, in fact, been
eliminated.

Anyway, here we are again confronted with instability of the solution: a
very small perturbation of this {\em pdf}, obtained by adding an
arbitrary ``infinitesimal" function with zero total integral will
resolve the ambiguity in a way which depends completely on the exact
form of the perturbation, {\em however small} its size. In this case it
is not even necessary to consider narrow spikes as in previous examples:
the instability can be obtained with perfectly smooth and slow--varying
functions.

Also, there is  no obvious way to extend to this case the modifications
suggested in \cite{Roe-Wood} for the Poisson with background example.

\subsubsection{Indifferent distributions}

In order to better illuminate the nature of the problem that is
frustrating the attempts at obtaining sound classical limits it is
useful to examine a ``trivial" example: a probability distribution that
does not depend on the value of $\mu$:

\begin{equation} p(x|\mu) = p(x) \end{equation}

For simplicity, consider the specific case of a distribution of a
discrete observable with just two values (`A' and `B') depending on a
parameter with just two possible values (`P' and `Q'), given by the
following table:

\begin{equation}\nonumber
\begin{tabular}{l||c|c||} &	P		&Q\\
\tableline	\tableline A		& 0.95&	0.95\\ \tableline B		& 0.05&	0.05\\
\tableline	\tableline \end{tabular}
\end{equation}

Clearly in this case the observable is not providing any information on
the parameter. What is a ``sensible" band in this case? Obviously no
conclusion can be drawn, so it should be clear that the only acceptable
band is the one that includes the whole table.  On the other end, most
rules will yield an empty region in case `B'.

The LR is constant everywhere, so the LR--ordering allows you to choose
any Neyman band. In force of the economical principle that unneeded
overcoverage is to be avoided, the best solution appears to be the band
that covers only the upper row of the table, and leaves an empty region
for case `B', just as the Crow rule.

In principle, nothing forbids to even choose arbitrarily to reject one
of the two values `P' and `Q' and keep the other in the case `B' is
observed, thus accepting some overcoverage. That choice is {\em very}
unreasonable from a physicist's viewpoint: it means one can conclude
essentially anything from the occurrence of event `B'. For instance,
when investigating the neutrino mass, one can make an ``experiment" by
doing something completely unrelated, for instance, by throwing a pair
of dice. Since the probability of getting, say, 6 on both dice is
$<3\%$, if that event actually occurs, one is entitled to exclude a mass
range of his choice at 97\% CL. I think very few persons would accept
this as a sensible inference, yet the procedure is perfectly correct
from the point of view of Neyman's definition, and is compatible with
LR--ordering, too.

Here the criteria of coverage shows clearly its inadequacy: to obtain a
sensible answer it is not enough that no more than 5\% of the outcomes
are excluded for every $\mu$, it would also be necessary to make sure in
some way that the choice one makes is not based on information
irrelevant for distinguishing different values of the parameter.

It should be clear at this point that this is the fundamental weakness
of Neyman's definition (\ref{eq:Neyman}), from which all problems arise.
As for the LR ordering rule, it appears to be going somehow in the right
direction, but it is unable to provide a clear--cut answer to a simple
problem like this.

Things get even worse if a small perturbation of the indifferent band is
introduced, leading to the following situation:

$$ \label{tab:indiffpert} \begin{tabular}{l||c|c||} & P & Q\\ \tableline
\tableline A & $0.95+\epsilon$ & $0.95-\epsilon$\\ \tableline B &
$0.05-\epsilon$ & $0.05+\epsilon$\\ \tableline	\tableline \end{tabular}
$$

Common sense clearly suggests not to draw any conclusion in this case,
too (not at 95\% CL, at least).

The LR method instead provides now unambiguously the answer of a
confidence region covering all but the lower left cell. This means, no
conclusion is drawn from observing event `A', but `P' is excluded if
event `B' is observed.

Admittedly, now `Q' is the maximum Likelihood estimation of the
parameter , but the difference with the previous case of ``crazy
inferences" is infinitesimal. When we claim that the conclusion has 95\%
CL, what meaning can we attach to this number if, however small the
difference, the CL is always 95\% ? It looks like a too strong statement
for an infinitesimal difference between the two hypothesis. Note that
the band obtained for this case is exactly the same that would have been
obtained from the following distribution, at the same CL:

$$ \label{tab:indiffpert2} \begin{tabular}{l||c|c||} & P & Q\\
\tableline \tableline A & $0.95+\epsilon$ & $0.05+\epsilon$\\ \tableline
B & $0.05-\epsilon$ & $0.95-\epsilon$\\ \tableline	\tableline
\end{tabular} $$ yet the two situations are intuitively very different
in terms of sensitivity of the experiment to the value of the parameter.

This example sheds serious doubts on the meaningfulness of valutations
of the sensitivity of a designed experiment based on expected confidence
limits calculated with any current rule. Again, this is a very serious
inconvenient, and the failure in handling a so simple example should
make us suspicious of many other bands, or maybe of {\em all} Neyman's
confidence bands.

\section{Proposal of a classical solution}

\subsection{Nature of the problem}\label{sec:analysis}

All proposed classical rules for building confidence bands meet with
severe difficulties when confronted even with simple problems. This is
true also for the recently proposed LR-ordering which appears to do only
slightly better that older recipes.

It is worth noting that the characteristics of the most common pdf's
taken as example of the difficulties (first two of previous section) has
often lead to speak of a ``problem of bounded regions" or of  ``small
signals". However, the additional examples provided should be sufficient
to clarify that the presence of a boundary, or the smallness of the
number of counts are just accidents without connection to the root of
the problem.

One should ask what is the exact reason for considering the previous
examples of confidence limits unacceptable. Their results are obviously
mathematically correct. The problem is not of 'mathematical' or
'statistical' but of 'physics' nature: one is lead to setting confidence
limits which are intuitively 'unpleasant' to the physicist, sometimes
even paradoxical. We don't want to accept a result like and empty
confidence region, which we {\em know} is false no matter what $\mu$ is,
because we feel we could do better inferences by keeping that fact into
account somehow. Indeed a result of this kind does not convey much
useful information to the reader. The same can be said for the softer
pathology of  ``unreasonably small" confidence regions.

It is hard to avoid the suspect that problems of the kind exemplified
above might be happening even in other cases that we usually regard as
problem--free, just because the problem is not so apparent to intuition.

Each of the encountered problems lies in the choice of a particular
confidence band, and in principle can be cured by simply choosing a
different band. However, one cannot content oneself with avoiding the
problem case by case by rejecting unreasonable results ``by hands". The
above described weakness are so important to undermine the physicist's 
belief in the meaning of CL. It is therefore necessary to find a general
way to avoid any such "unwanted" conclusion, even in possibly softer,
hidden forms.

The question is: can we state precisely what properties we require from
a confidence band to call it 'physically sensible' ? Does a single
well-defined procedure exist to construct one in a generic case ?

Let us look in more detail at the meaning of confidence limits.

The Neyman's definition of CL can be phrased in the following way:

``An algorithm is said to have Confidence Level CL if it provides {\em
correct} answers with probability {\em at least} CL, whatever the value
of $\mu$ (or whatever its probability distribution, if it has one)"

From a practical point of view, this means that if one considers, for
instance, the set of all published limits at 95\% CL, the expected
fraction of them which is indeed ``wrong" (that, is, the limits do not
include the true value) is 5\% (or smaller, if there is some
overcoverage).

For comparison, the definition of Bayesian credibility level, when
phrased in a similar way, sounds like:  ``An algorithm is said to have a
Bayesian credibility level BL if all answers it produces have at least a
probability BL of being correct,  provided $\mu$ has the (known)
probability distribution $\pi(\mu)$". In exchange for an additional
assumption (the a--priori distribution) the Bayesian method provides a
probability statement about {\em each} measurement.

The classical approach cannot possibly do that, since the concept of
probability for a single result of being correct simply cannot be
formulated in the classical language: each particular result is {\em
either} true {\em or} false, since the unknown parameter is taken to
have {\em one} (if unknown) value, rather than a distribution of
possible values. Superficially, however, the classical method seems to
provide a close performance, when saying that the whole set of results
contains only a limited fraction of wrong results ($<1-CL$).

There is, however, a subtle difference between a statement {\em
extracted from a sample} containing 95\% correct statements, and a
statement that {\em has} a 95\% probability of being true. The
difference is that in the first case some manifestly false or very
unlikely statement are allowed to be part of the set (e.g. , empty
confidence regions), provided they are a minority, while in the second
case this is not possible: every single possible Bayesian inferred
result is forced to be as likely as all others.

This is the fundamental reason for the absence of pathological
conclusion in the Bayesian approach, that keeps tempting the classical
physicist. Its appeal is so strong that even the purest classical papers
show some slight inclination toward Bayesianism.

As an example, the method suggested in \cite{F-C} for evaluating an
experiment sensitivity uses the concept of ``average limit", that in
general requires a {\em a--priori distribution} of the parameter to be
assumed, even if the paper only consider the special case of no signal
for that purpose. In \cite{Roe-Wood}, after a nice classical suggestion
for solving the Poisson problem classically, the results are compared to
Bayesian results and their similarity is taken as a support to their
soundness, notwithstanding the fact that if one had to change the
a-priori distribution to something different the Bayesian result will
change completely, while the classical result will always stay the same.

It becomes therefore imperative to ask the question: is there any way to
give to the classical method the same solidity without introducing any
Bayesian element ? If there is none, than it may be simpler to abandon
the classical method completely and use Bayesian concepts instead.

The purpose of this paper is to suggest that there is indeed a way to
obtain the desired properties in the classical framework.

The definition of CL ensures that the result will be correct at least a
fraction CL of the cases. An empty region is never a correct conclusion,
because $\mu$ has {\em some} value by hypothesis. The definition of CL 
is not meant to prevent wrong conclusions: it just makes sure that they
happen rarely, the only limitation to empty CR being that it must not
occur with probability greater than 1-CL, whatever the value of $\mu$.
In fact, it is easy to see that, given a set of values of $x$ that has
total probability $<1-CL$ for any $\mu$,  it is always possible to
assign the empty set as confidence region for all $x$ in the set,
provided the rest of the band is properly adjusted.

This fact may even appear as a kind of inescapable ``law" of classical
statistics. After all, in formal logic one has that given a
contradictory (impossible) assumption one can rigorously derive any
statement. We might have to accept a kind of probabilistic analogue as
well, that is, from the occurrence of an {\em unlikely} event one can
{\em statistically infer} any statement.

However, this is actually not the case. What disturbs the physicist is
not the mere possibility of getting wrong results, which he obviously
has to accept, but that one might get a wrong result {\em and know it}.
One could say that those are ``unlucky" experimental results. But there
are good reasons to refuse to surrender to the occurrence of ``unlucky"
results: common sense suggests that once we get a result that we know to
be uncommon, there should be a way to correctly account for its rarity,
rather than getting confused by it.

In a way, what we really need is to make sure that all experimental
outcomes get uniform treatment, like in the Bayesian method.

In this respect, in is worth noting that the strength of the definition
of CL lies in its invariance for any transformation of the space of
parameters $\mu$, even non-continuous. That is, all points of the
parameters space get the same treatment, the metric and even the
topology of the parameter space being irrelevant. We could even say that
this is the {\em essence} of the classical statistics. This is to be
contrasted with the Bayesian approach, where the a-priori distribution
sets a well-defined metric in the parameter space. Then, why it happens
that the classical methods seem to be so much worse than the Bayesian in
assuring invariance in the {\em outcome} space ?

As a matter of fact, Neyman's definition of CL (\ref{eq:Neyman}) is
symmetric for all values of $x$. However, most rules for constructing
confidence bands break this symmetry: it is easy to see that by
performing a change of variable in $x$ one obtains different
bands\footnote{LR ordering is an important exception. See sec.\
\ref{sec:properties} for a discussion of this point}. This allows the
mere fact that a particular experimental outcome is unlikely for some
parameter value to be used to exclude that value, regardless to the fact
that the outcome might be unlikely for entirely different reasons than
the value of the parameter being sought. That probability might be low
for {\em every} value of the parameters, so the exclusion of that
particular value of $\mu$ is taken on the basis of {\em irrelevant}
information. Neyman's construction, while compatible with total symmetry
in $x$, does not explicitly enforces it, because it applies
independently to each value of $\mu$, and there is no way to tell
whether the distribution of $x$ has any dependence on the value of
$\mu$.

We need therefore to find a way to prevent the introduction of
information irrelevant to the determination of the parameters in the
choice of the confidence band. It should be intuitively clear that there is a
connection between the contamination from irrelevant information and the
unequal treatment of various possible experimental outcomes that is the
basis of paradoxical results.

The present approach is in some way the opposite of the attempts to
improve the classical method by the addition of Bayesian elements: it
goes in the direction of an even stricter classical orthodoxy. The use
of any metric or topological property of the $x$ space is regarded as an
``a priori bias" producing unequal treatment of some values. That is a
kind of contamination from ``Bayesianism" that needs to be eradicated
from a pure classical method, which ought to use only the
information contained in the {\em pdf}.

\subsection{A stronger concept of Confidence}\label{sec:proposal}

We formalize the request that the choice of Confidence Regions must not
be based on irrelevant information in the following requirement.

Suppose we take a subset of $x$ values and rescale all likelihoods
$p(x|\mu)$ by the same arbitrary factor, (we have to re-normalize the
pdf for the rest of $x$ values after that, of course). A physically
sensible rule for constructing confidence bands must be invariant under
this kind of transformation, since the overall absolute level of
probability of the events $x$ does not affect the information that can
be obtained on $\mu$. More precisely, we want to restrict the set of all
possible confidence bands to a subset that satisfies the following
property, which will be called local scale invariance:

DEFINITION -- Let $x \in X$ be an observable and $\mu \in M$ a
parameter. Let ${\cal R}$ be a rule for selecting confidence bands, that
is, a function that associates to each possible distribution $p(x|\mu)$
a set of Neyman confidence bands with a given CL. We say that ${\cal R}$
is a {\it locally scale--invariant} rule if for any two pdf's $p(x|\mu)$
and $p'(x|\mu)$ such that $p'(x|\mu) = c \cdot p(x|\mu)$ for all $\mu
\in M$ and for all $x \in \chi \subset X$ (with $c$ positive constant),
and for every confidence band $B\in{\cal R}(p)$, there exist a band
$B'\in{\cal R}(p')$ such that $B(x) = B'(x)$ for every $x \in \chi$.

This requirement is simple, general, and intuitively satisfying: it says
that whatever algorithm we want to use to choose a CR for a certain set
of possible observations, it must not be influenced by anything else
than the dependence on $\mu$ of the probability of the observations in
question. Note that both the observable and the parameter space can be
completely generic sets. We keep requiring all bands to comply with
Neyman's condition, which however does not by itself guarantee the above
property, neither it does any of the proposed algorithms for producing
confidence bands, including the LR-ordering. The latter appears clearly
from our previous discussion of the example of Poisson with background.

It is interesting to observe that the {\em rank} assigned to $x$ by the
LR--ordering rule is indeed invariant under the above transformation,
but the {\em coverage} criteria used to decide when to stop adding
values of $x$ to the acceptance region is not. The normalization
constant creates the difficulty here, since one can have a region
rejected in one case, that cannot be rejected in the other because its
contribution to the total integral may be too large.

We will now show that this seemingly weak requirement is actually very
stringent in determining the set of allowed confidence bands, and that
it can be turned into a well definite procedure for constructing
bands.

This is seen from the following theorem.

THEOREM --  The largest set of locally scale--invariant bands coincides
with the set of bands satisfying the following requirement:

For every $\mu \in M$ and every $\chi \subset X$:

\begin{equation}\label{eq:loc_cond} {p(x \in \chi \and \mu \not \in
B(x)|\mu) \over \sup_\mu \ p(x \in \chi|\mu)} \leq 1-CL. \end{equation}

whenever the denominator is non--zero.

PROOF:
 
Part 1 -- All bands in a locally scale--invariant set satisfy condition
(\ref{eq:loc_cond}).

Suppose (\ref{eq:loc_cond}) does not hold. Then there is a band $B$, a
subset $\chi$ and a parameter value $\bar\mu$ such that

\begin{equation}\label{eq:no_loc} {p(x \in \chi \and \bar\mu \not \in
B(x)|\bar\mu) \over \sup_\mu \ p(x \in \chi|\mu)} > 1-CL$$
\end{equation}

Then consider a new pdf defined inside $\chi$ by: $$p'(x|\mu)={p(x|\mu)
\over \sup_\mu \ p(x \in \chi|\mu)}$$ and arbitrarily extended outside
$\chi$. This is always possible since by construction $\int_{\chi}
p'(x|\mu) dx \leq 1$ for every $\mu$.

Obviously, for every $\mu$: $$p'(\mu\not\in B(x)|\mu) \geq p'(x \in \chi
\and \mu\not\in B(x)|\mu)$$ And from (\ref{eq:no_loc}): $$p'(x \in \chi
\and \mu\not\in B(x)|\mu) = \\ {p(x \in \chi \and \mu\not\in B(x)|\mu)
\over \sup_\mu \ p(x \in \chi|\mu)}> 1-CL$$ then $$p'(\bar\mu\not\in
B(x)|\bar\mu) > 1-CL$$ which contradicts Neyman's condition. Therefore
$B$ could not be part of an invariant set, in contradiction with the
hypothesis. Therefore eq. (\ref{eq:loc_cond}) is proved.

Part 2 -- The set of all bands satisfying (\ref{eq:loc_cond}) is a
locally scale--invariant rule.

First of all, note that (\ref{eq:loc_cond}) implies Neyman's condition
as a special case (just take $\chi=X$).

Take any $p$, $B$, $\chi \in X$, $c>0$, and $p'=c\cdot p$ for all $x\in
\chi$. Note that the ratio in (\ref{eq:loc_cond}) does not change when
the pdf is scaled by a constant, so if B satisfies (\ref{eq:loc_cond})
for $p$ in $\chi$ and all its subsets it will also satisfy it for $p'$.
Let's define $B'=B$ in $\chi$ and $B'=M$ (the whole parameter space)
outside $\chi$. Then, for any $\xi\subset X$ we have: \begin{eqnarray*}
{p'(x \in \xi \and \mu \not \in B'(x)|\mu) \over \sup_\mu \ p'(x \in
\chi|\mu)}=\\ ={p'(x \in (\xi\cap\chi) \and \mu \not \in B'(x)|\mu)
\over \sup_\mu \ p'(x \in \chi|\mu)}\leq\\ \leq {p'(x \in (\xi\cap\chi)
\and \mu \not \in B'(x)|\mu) \over \sup_\mu \ p'(x \in
(\xi\cap\chi)|\mu)}=\\ = {p(x \in (\xi\cap\chi) \and \mu \not \in
B(x)|\mu) \over \sup_\mu \ p(x \in (\xi\cap\chi)|\mu)}\leq 1-CL
\end{eqnarray*} This means $B'$ satisfies (\ref{eq:loc_cond}) for the
distribution $p'$. Since we defined $B'=B$ in $\chi$, this proves local
scale--invariance of the set of bands given by (\ref{eq:loc_cond}).

Part 1 and 2 together show that the two sets coincide, concluding the
proof. Note that they implicitly prove that the ``largest set of
locally scale--invariant bands" indeed exists, which was not granted {\em
a priori}\footnote{We could have proved the existence beforehand, by
observing that the union of any number of invariant rules is still an
invariant rule, therefore the largest invariant rule is immediately
identified as the union of all possible invariant rules.}.

Condition (\ref{eq:loc_cond}) is clearly connected with the intuitive
concept of uniform treatment of all experimental results, and offers a
much clearer indication than the equivalent scale--invariance
requirement about how to construct in practice a satisfying confidence
band.

It also appears as a natural extension of Neyman's $CL$ concept, because
it amounts to simply applying at local level the same requirement Neyman
imposed on the observable space as a whole.

This fact suggest an alternative formulation: rather than regarding the
(\ref{eq:loc_cond}) as a rule for identifying a particular subset of
confidence bands, we can take this condition as a new, more restrictive,
definition of limits within the classical framework (``Strong Confidence
Limits") and define a new quantity (``strong CL", or ``sCL") in analogy
with the usual $CL$ (eq.\ (\ref{eq:Neyman})):

\begin{equation}\label{eq:sCL_def}
sCL(B) =  1 - \sup_{\mu,\chi} {p(x \in \chi \and\mu\not\in B(x)|\mu) \over
\sup_\mu \ p(x \in \chi|\mu)} \end{equation}

The strong CL is then a quantity that can be evaluated for a completely
arbitrary band, just as the regular CL. Note that it is always $sCL \leq
CL$, in accordance with the greater strength of the concept.

\section{Properties of Strong Confidence Regions}\label{sec:properties}

The meaning of strong confidence can be summarized as follows: take a
subsample of possible experimental results, however defined. While it is
still not guaranteed that the probability for them to be correct is at
least CL as with Bayesian methods, what we gained over Neyman's CL is
that, independently of the a--priori distribution of $\mu$, the number
of wrong result is a small fraction of the {\em maximum expected} number
of results of that kind. That is, there may be distributions for $\mu$
that lead to {\em all} results in that category to be false, but in that
case those results will present themselves much more rarely than when
they lead to correct conclusion, and this holds for all possible results
in the same way. This is basically how far we can go within the
classical framework in terms of getting ``individually certified"
results\footnote{Note that, even if one assumes that a distribution of
the parameter {\em exists}, a probability statement about each result is
impossible to obtain in the classical framework without {\em knowing}
the distribution (a priori), unless one chooses the trivial solution of
the band covering the whole space.}.

The possibility of empty confidence regions is ruled out here in full
generality, unlike the case of LR ordering: if there is a set $\chi$ of
values for which the confidence region is empty, then obviously there
exist a $\mu$ for which the ratio on left side of (\ref{eq:loc_cond}) is
arbitrarily close to 1. Unless CL=0, that means the total probability of
$\chi$ is identically zero for all $\mu$.

It is also easy to see that strong bands are stable for small
perturbations of the pdf like those previously discussed in the
examples. This is due to the fact that the requirements being made are
based on integrals of the pdf rather than its punctual values. The
integrals on all subsets that are not too small stay the same after the
addition of perturbations of small total probability. The effect is only
seen on small scales, and it just forces the addition of the small
region where the perturbation is large to the unperturbed band.

\subsection{Independence from change of variable}

The above defined strong bands have another interesting property: they
are invariant under any change of variable in $x$--space. This is
obvious since the probabilities appearing in the ratio in
(\ref{eq:loc_cond}) scale proportionally under any change of variable.
We stressed before that the strength of the classical approach lies in
its independence from metric in $\mu$--space, that is, in its equanimity
with respect to every value of $\mu$, in contrast with the Bayesian
approach where all values are explicitly weighted for relative a-priori
importance.

It should be clear as well that the use of a particular metric of $x$
space in constructing a CR is a way to introduce a--priori
discriminations between values of the $x$, that is, to introduce
arbitrary (irrelevant) information in the choice of the confidence band,
so it is not surprising that this invariance is a consequence of our
approach.

Amongst all common rules for selecting CR, LR-ordering is the only one
to be independent from transformations of $x$. The partial success of
the LR ordering principle might in the end be traced back to its
compliance with this requirement of independence from metric in $x$
space.  In fact, the LR ordering rule is equivalent to the {\em
narrowest} band in that particular metric in parameter space that makes
the maximum likelihood value constant for all $\mu$. 

We have seen, however, that while this property is desirable, and
probably necessary for physically sensible results, it is not sufficient
to ensure them.

\subsection{Construction of Strong Confidence Regions}
\label{sec:construction}

A simple and useful corollary of (\ref{eq:loc_cond}) is:

COROLLARY -- If the observable is discrete\footnote{It does not hold in
full generality for a continuous variable, since it is always possible
to choose an arbitrary band for single isolated $x$ without affecting
any of the formulas above, that always refer to events or set of events
with of non-zero total probability. However, it is intuitively expected
that it holds for continuous variables too, provided some regularity
condition is asked}, than for every value of $x$, any strong band always
includes all values of $\mu$ such that $p(x|\mu) > p(x|\hat
\mu)\cdot(1-sCL)$.

PROOF: just put $\chi = \{x\}$ in (\ref{eq:loc_cond}).

This immediately shows another appealing feature of strong CL:  it is
forbidden to exclude any value of $\mu$ having a likelihood ``close" to
the max-likelihood value. Again, this is not a property of any other
method (LR--ordering just tends to give such values somewhat high ranks, and
$\hat \mu$ gets the highest rank whenever it exists, but there is no guarantee on their
actual inclusion in the band. Other rules do even less than that).

Note that, just as in Neyman's CL, in a generic case there may be many
different ``legal" bands for a  given pdf and sCL, therefore the
question of the choice between them reappears. However, since there is
now no fear of unreasonable results, the only reason for pinpointing a
general and unique choice is just to prevent the possible distorted
practice of choosing the band after the experiment, and the question is
largely a matter of convenience. In order to be coherent with the
spirit of the current approach, however, the choice must be formulated
in such a way to be invariant under any transformation of $x$.

For instance, a good choice might be to minimize the {\em coverage} for
every value of $\mu$ independently. This makes for the lowest possible
CL for the given sCL. Conversely, the bands chosen in this way will have
the highest sCL for a given CL. They can be considered with good reason
the ``best band" for a given CL, in case an experimenter wishes to fix
the desired value of CL as usual, rather than the sCL. Obviously, if
the maximum sCL corresponding to a given CL is small or even zero, that
means no physically sensible band is possible without increasing the CL
(that is,``overcovering" of all values of $\mu$ is necessary).

In practice the freedom of choice is often very limited, since the
``core region" identified by the above corollary must be completely
included by any legal strong band at the given sCL. That core region is
defined only by the the pdf for the {\em local} values of $x$, therefore
is not affected by changes in the pdf for other values.

The actual determination of the bands in other than the simplest cases
requires numerical calculations. We now describe a simple algorithm to
construct in practice a band satisfying the criteria.

In order to do numerical calculations, the pdf must be discretized if
the parameter or the observable are continuous. This is achieved by
sampling the parameter space with an N--dimensional grid, and splitting
the space $X$ of the observable into a finite number of regions. Those
regions are considered as possible discrete outcomes, and their
probabilities are obtained by integrating the density $p(x|\mu)$ over
each of the regions. In this way, a rectangular matrix is obtained,
independently on the dimensionality of the $x$ and $\mu$ spaces, which
may be both arbitrary-length vectors of numbers. This matrix is used as
input in the following simple algorithm. 

All intervals of $x$ are initially assigned to the rejected region, that
is, the band is initialized to be empty. For each value of $\mu$, one
loops over all possible sets composed of any number of the chosen $x$
regions. The condition (\ref{eq:loc_cond}) is checked on all sets in
turn, and if found invalid, one of the regions in the current set is
added to the confidence band, and removed from any further checks. The
set of accepted regions obtained upon completion of this procedure for
all values of $\mu$ is a strong band.
The freedom in the choice of the region to be added to the band is what
allows different solutions to be generated. 

It is not obvious how to achieve the minimal coverage requirement
suggested above within this stepwise procedure. There are however simple
and reasonable recipes for performing the choice step of the algorithm. One
can, for instance, systematically choose the lowest/highest $x$ to get
the analogue of lower/upper limits in the standard approach, or choose
the region with the highest value of the ratio tested by condition
(\ref{eq:loc_cond}). The latter appears particularly natural and
has the interesting characteristics of representing an extension of the
LR--ordering rule to the sCL context, even if the result might be slightly dependent
on the order in which the sets of regions are being checked by the algorithm.

\subsection{Sample Applications}\label{sec:applications}

The definition of strong CL gives satisfying answers to all problems
listed in Sec.\ \ref{sec:examples}.

In some cases the solution follows immediately from the corollary above.

One of them is the ``indifferent"  pdf, where the conclusion that no
value of the parameter can be excluded, whatever the required sCL is
immediately found, and it is  stable for small perturbations of the pdf.

For the uniform distribution, the full range of $\mu$ for which
$L(\mu)>0$ gets included, whatever the chosen $sCL$. This strong
statement reflects the intuitive arbitrariness of any choice wishing to
exclude some value of a parameter in favor of others having exactly the
same likelihood. In fact, when a problem with uniform pdf is
encountered, most physicists don't even formulate a question of
Confidence Limits, but just quote the absolute extrema of the allowed
interval for $\mu$.

For the Poisson with background, it is easy to see that the result for
the case of zero observed events will be independent of background. The
probability of zero events is $e^{-\mu}e^{-b}$, so by changing the
expected background $b$ one changes the likelihood by a simple
multiplicative constant. From the definition of local scale invariance
one has immediately that the limits for this case cannot depend on $b$.
This statement needs a bit of clarification: we have remarked that the
strong band is not uniquely identified in a general case, therefore one
can make various choices. What is guaranteed here is that all possible
choices for the limits from zero counts for a given value of $b$ are
also acceptable choices for any other value of $b$. This does not imply
that one must necessarily make the same choice in the two cases.

We have calculated the confidence limits in the special case of $b$=3.0
using the simple method outlined in the previous section, and compared
the results with other classical methods in Table\ \ref{tab:Poisson}.
The upper, lower and the LR-ordering analogue choices mentioned above are
shown. The intervals obtained are wider than with any other method.

This should not  be considered a loss of power , but rather
regarded as a reflection of the higher standards of quality required to
the result. The parts of the band that would be excluded by other
methods are here included just on the same basis that yields the correct
conclusion for the zero--count case, and prevents crazy conclusions from
indifferent distribution: their likelihood is not low enough with
respect to the maximum value. These considerations suggest that one
should not consider this widening of the band a loss of power unless one
also considers a loss of power the inability to draw conclusions on the
mass of neutrinos by throwing dice.

\section{Summary}

The current methods for determining classical Confidence Limits produce
counter--intuitive results in a variety of situations. This includes the
recent proposals based on Likelihood Ratio ordering, that is not immune
from the problem of empty confidence regions.

By imposing the requirement that only the information contained in the
shape of the Likelihood function be used in determining the limits, a stronger
definition of classical limits is derived, which is a natural extension
of the original Neyman's condition.

This ``strong confidence limits" turns out to be immune to the problem
of empty accepted regions, and stable for small perturbations of the
probability distribution, at the price of some widening of the usual
limits.

\acknowledgments

I wish to thank Luciano Ristori of Istituto Nazionale di Fisica Nucleare
in Pisa for useful discussions and comments on the manuscript.

\appendix

\widetext \begin{table}[tbp] \caption{Comparison of
90\% confidence intervals for Poisson with background level of 3.0, for
LR ordering, modified LR\protect\cite{Roe-Wood}, and strong CL (``pseudo-LR", low, and high
band)}\label{tab:Poisson} \begin{tabular}{ddddddddddd} &LR& &LR modified& & sCL & &sCL low&
&sCL high& \\ $n$ (observed)& min&max&min&max&min&max&min&max&min&max\\
\tableline 0&0.0&1.08&0.0&2.42&0.0&2.8&0.0&2.4&0.0&3.1\\
1&0.0&1.88&0.0&2.94&0.0&3.6&0.0&3.0&0.0&4.5\\
2&0.0&3.04&0.0&3.74&0.0&4.5&0.0&4.0&0.0&5.9\\
3&0.0&4.42&0.0&4.78&0.0&5.9&0.0&5.4&0.0&7.5\\
4&0.0&5.60&0.0&6.00&0.0&7.5&0.0&7.0&0.0&9.7\\
5&0.0&6.99&0.0&7.26&0.0&9.4&0.0&8.5&0.0&11.2\\
6&0.15&8.47&0.42&8.40&0.0&10.9&0.0&9.9&0.0&12.8\\
7&0.89&9.53&0.96&9.56&0.0&12.4&0.0&11.4&0.0&13.8\\
8&1.51&11.0&1.52&11.0&0.4&13.8&0.4&12.7&0.4&15.2\\
9&1.88&12.3&1.88&12.22&1.0&15.2&1.0&14.1&1.0&16.8\\
10&2.63&13.5&2.64&13.46&1.7&16.6&1.7&15.5&1.7&18.3\\ \end{tabular}
\end{table} \twocolumn \end{document}